\documentclass[nofootinbib,superscriptaddress,longbibliography,prd,twocolumn,amsmath,amssymb,preprintnumbers]{revtex4-2}
\usepackage{bm}
\usepackage{graphicx}
\usepackage{hyperref}
\usepackage{color}

\newcommand{\bfe}{\mathbf{e}}
\newcommand{\bfP}{\mathbf{P}}
\newcommand{\bfN}{\mathbf{N}}
\newcommand{\bfF}{\mathbf{F}}
\newcommand{\bfH}{\mathbf{H}}

\newcommand{\tP}{\tilde{P}}
\newcommand{\tbP}{\tilde{\bfP}}
\newcommand{\p}[1]{\protect{(#1)}}

\newcommand{\us}{u^{(\mathrm{s})}}
\newcommand{\pfrac}[2]{\left(\frac{#1}{#2}\right)}

\newcommand{\rmd}{\mathrm{d}}

\newcommand{\Tr}{\mathrm{Tr}}
\newcommand{\Gf}{G_\mathrm{F}}

\newcommand{\figscale}{0.7}

\newcommand{\avgP}{\overline{P}}

\newcommand{\nugas}{\texttt{NuGas}}

\newcommand*{\UNM}{Department of Physics \& Astronomy, University of New Mexico, Albuquerque, New Mexico 87131, USA}

\newcommand*{\LANL}{Theoretical Division, Los Alamos National Laboratory, Los Alamos, New Mexico 87545, USA}

\begin{document}

\title{Flavor isospin waves in one-dimensional axisymmetric neutrino gases}

\author{Huaiyu Duan}
\email{duan@unm.edu}
\affiliation{\UNM}

\author{Joshua D. Martin}
\affiliation{\LANL}

\author{Sivaprasad Omanakuttan}
\affiliation{\UNM}

\begin{abstract}
    Flavor oscillations can occur on very short spatial and temporal scales in the dense neutrino media in a core-collapse supernova (CCSN) or binary neutron star merger (BNSM) event. Although the dispersion relations (DRs) of the fast neutrino oscillations can be obtained by linearizing the equations of motion (EoM) before the emergence of any significant flavor conversion, one largely depends on numerical calculations to understand this interesting phenomenon in the nonlinear regime. In this work we demonstrate that there exist nontrivial solutions to the flavor EoM that govern the fast oscillations in one-dimensional axisymmetric neutrino gases. These solutions represent a coherent flavor isospin wave similar to the magnetic spin wave in a lattice of magnetic dipoles. We also compute the DRs of such waves in some example cases which are closely related to the DRs of the fast neutrino oscillations obtained in the linear regime. This result sheds new light on the long-term behavior of fast neutrino oscillations which can have various implications for the CCSN and BNSM events. 
\end{abstract}

\date{\today}

\preprint{LA-UR-21-29708}

\maketitle

\section{Introduction}\label{sec:intro}

Copious neutrinos are emitted during a core-collapse supernova or binary neutron star merger event which help to shape the physical and chemical evolutions of these fascinating astrophysical systems. An intriguing question that has long been studied is what roles neutrino oscillations play in such events. Because neutrino-neutrino coherent scattering can exist even in dense, free-streaming neutrino gases \cite{Fuller:1987aa, Notzold:1987ik, Pantaleone:1992xh}, the dense neutrino media near a compact object can experience flavor transformation collectively (see, e.g.\ Refs.~\cite{Duan:2010bg, Chakraborty:2016yeg} for reviews and references). This collective oscillation of the neutrinos can occur on very short distance scales, and thus dubbed as ``fast oscillations'', when the angular distributions of the neutrinos are different \cite{Sawyer:2005jk, Sawyer:2015dsa, Chakraborty:2016lct}. (See, e.g., Ref.~\cite{Tamborra:2020cul} for a review.)

At the forefront of these recent developments is the research on the fast flavor oscillations in time-dependent, one-dimensional (1D) neutrino gases. Although the flavor evolution of this model can be inferred from the neutrino electronic lepton number (ELN) distribution and the dispersion relation (DR) when little flavor conversion has occurred \cite{Izaguirre:2016gsx, Capozzi:2017gqd, Yi:2019hrp}, the few numerical explorations that have been made public so far seem to disagree qualitatively in the nonlinear regime. While our previous works show that fast oscillations can be coherent in space and time \cite{Martin:2019gxb,Martin:2021xyl}, the calculations from two other groups suggest that flavor depolarization should occur instead \cite{Bhattacharyya:2020jpj,Bhattacharyya:2020dhu, Richers:2021nbx}. A recent study shows both the coherent flavor evolution in the short term and flavor depolarization in the long term when neutrinos are allowed to wrap around the periodic box \cite{Wu:2021uvt}.

In this work, we try to gain further understanding of the fast oscillation phenomenon by studying the semi-analytic solutions to the equations of motion (EoM) that govern the neutrino flavor oscillations in the nonlinear regime. The idea of this work is rooted in the gyroscopic pendulum picture that explains the two-flavor, bipolar oscillations of a homogeneous and isotropic neutrino gas which initially consists of pure $\nu_e$'s and $\bar\nu_e$'s of the same energy \cite{Hannestad:2006nj}. With a suitable initial condition, a gyroscopic pendulum can precess smoothly without any wobbling. This corresponds to the pure precession of the flavor Bloch/polarization vectors $\bfP$ \cite{Sigl:1993ctk} or the flavor isospins of the neutrinos \cite{Duan:2005cp} in the flavor space. For the inverted neutrino mass ordering and at sufficiently large neutrino densities, the precession solution describes a flavor pendulum in the sleeping top regime where it stays in the upright configuration or the north pole. As the neutrino densities decrease, the precession of the gyroscopic pendulum at the north pole becomes unstable, and the stable precession occurs at lower latitude instead \cite{Duan:2007mv}. Similar pure precession solutions have been found for the neutrino gases with continuous energy spectra \cite{Raffelt:2007cb,Raffelt:2007xt} and are used to explain the spectral swap/split phenomenon in the supernova neutrino calculations \cite{Duan:2006jv,Duan:2006an}. Similar solutions have also been found for homogeneous but anisotropic neutrino gases \cite{Xiong:2021dex}. 

It has been predicted that a precession solution to the flavor EoM of an inhomogeneous neutrino gas describes a flavor isospin wave similar to the magnetic spin wave in a lattice of magnetic dipoles \cite{Duan:2008fd}. In this paper, we show that the fast flavor isospin waves of the following form can indeed exist in 1D axisymmetric neutrino gases:
\begin{subequations}
    \label{eq:ansatz}
    \begin{align}
        P_{u,1}(t,z) &= \tP_{u,\perp}\cos(\Omega t - K z + \phi), \\
        P_{u,2}(t,z) &= \tP_{u,\perp}\sin(\Omega t - K z + \phi), \\
        P_{u,3}(t,z) &= \tP_{u,3},
    \end{align}        
\end{subequations}
where $P_{u,i}(t,z)$ ($i=1,2,3$) are the three components of the flavor Bloch vector $\bfP_{u}(t,z)$ of the neutrino with the $z$ velocity component $v_z=u$ and at time $t$ and location $z$, $\tP_{u,\perp}$ and $\tP_{u,3}$ are two real constants that satisfy the normalization condition $\tP_{u,\perp}^2 + \tP_{u,3}^2=1$, and $\phi$, $\Omega$ and $K$ are the initial phase, frequency and wave number of the flavor isospin wave, respectively. In fact, the DRs of the fast neutrino oscillations proposed in Ref.~\cite{Izaguirre:2016gsx} are those of the wave solutions in the limit $P_{u,3}\rightarrow1$. Because the flavor configuration with $P_{u,3}=1$ satisfies the flavor EoM trivially, we will call such solutions trivial wave solutions in the rest of the paper. The DR branches of trivial wave solutions with real $K$ and $\Omega$ correspond to the flavor pendulum in the sleeping top regime, and those with real $K$ and complex $\Omega$ are like the pendulum in the upright configuration but outside the sleeping top regime. Naturally, one expects that there should exist nontrivial flavor isospin waves with $P_{u,3}<1$ which correspond to the flavor pendulum precessing at a latitude lower than the north pole. We will show that this is indeed the case.

The rest of the paper is organized as follows. In Sec.~\ref{sec:eom} we write down the EoM that governs the fast neutrino oscillations and establish the formalism. In Sec.~\ref{sec:solution} we derive the self-consistent equations that can be used to determine the characteristic quantities of the flavor isospin waves. In Sec.~\ref{sec:examples} we showcase some of the wave solutions and demonstrate how they are related to fast neutrino oscillations. Finally, in Sec.~\ref{sec:conclusions} we discuss the potential physical significance of the flavor isospin waves and conclude.

\section{Equation of motion}\label{sec:eom}

We consider the flavor mixing between the electronic flavor ($e$) and the $\tau$ flavor in a dense 1D, axisymmetric neutrino gas, where the nominal $\tau$ flavor is an appropriate linear superposition of the physical $\mu$ and $\tau$ flavors. Assuming that the neutrino gas consists of nearly homogeneous neutrinos and antineutrinos in the $e$ and $\tau$ flavors in the beginning, one can define the following neutrino ELN distribution \cite{Izaguirre:2016gsx}:
\begin{align}
    G(u) = \frac{2\pi}{n_{\nu_e}^\p{0}} \int_0^\infty \Tr[(\rho^\p{0} - \bar\rho^\p{0})\sigma_3]\,\varepsilon^2\rmd\varepsilon,
\end{align}
where $n_{\nu_e}^\p{0}$ is the initial $\nu_e$ density, and $\rho^\p{0}$ and $\bar\rho^\p{0}$ are the initial flavor density matrices of the neutrino and antineutrino in the weak-interaction basis, respectively \cite{Sigl:1993ctk}, $\sigma_3$ is the third Pauli matrix, and the integral is performed over the neutrino energy $\varepsilon$. 
The EoM that governs the fast oscillations can be written as (see, e.g., \cite{Martin:2019gxb})
\begin{align}
    (\partial_t + u\partial_z) \bfP_{u} = 
    [\lambda \bfe_3 + \mu (\bfN - u\bfF)]\times\bfP_u,
    \label{eq:eom-lambda}
\end{align}
where the strength of the matter potential and the neutrino background in Eq.~\eqref{eq:eom-lambda} are defined as $\lambda=\sqrt2 \Gf n_e$ and $\mu=\sqrt2\Gf n_{\nu_e}^\p{0}$, respectively, with $n_e$ being the (net) electron number density, and 
\begin{align}
    \bfN &= \int_{-1}^1 G(u)\bfP_u\,\rmd u\\
    \intertext{and}
    \bfF &= \int_{-1}^1 u G(u)\bfP_u\,\rmd u
\end{align}
are the ELN density and flux (in the unit of $\mu$), respectively. In the weak-interaction basis [with basis vectors $\bfe_i$ ($i=1,2,3$)], the probability of finding the neutrino or antineutrino to be in the electron flavor is  
\begin{align}
    \mathcal{P} = \frac{1+\bfP\cdot\bfe_3}{2} = \frac{1 + P_{3}}{2}.
\end{align}
We have ignored the vacuum Hamiltonian in Eq.~\eqref{eq:eom-lambda} which does not affect the flavor dynamics on the scales of fast oscillations \cite{Sawyer:2015dsa,Izaguirre:2016gsx,Martin:2021xyl}. In this approximation, the flavor evolutions of all the neutrino and antineutrino fields with the same velocity $u$ but different energies are represented by the same Bloch vector $\bfP_u$. Following the convention in the fast oscillation literature, we will measure the time and distances in $\mu^{-1}$ by setting
\begin{align}
    \mu = 1.
\end{align}

If $n_e$ is uniform on the fast oscillation scale, Eq.~\eqref{eq:eom-lambda} can be further simplified in the reference frame that rotates about $\bfe_3$ with frequency $\lambda$ \cite{Duan:2005cp}. In this co-rotating frame,
\begin{align}
    (\partial_t + u\partial_z) \bfP_u
    = \bfH_u \times \bfP_u
    = (\bfN - u\bfF) \times \bfP_u.
    \label{eq:eom}
\end{align}
Integrating the above equation with weight $G(u)$, one obtains
\begin{align}
    \partial_t \bfN + \partial_z \bfF = 0.
    \label{eq:cons}
\end{align}
Along the flavor axis (in the direction of $\bfe_3$), this equation gives the conservation of the ELN:
\begin{align}
    \partial_t N_3 + \partial_z F_3 = 0.
\end{align}

\section{Wave solutions} \label{sec:solution}

The Bloch vectors that satisfy the wave ansatz in Eq.~\eqref{eq:ansatz} obey the following wave equation:
\begin{align}
   (\partial_t + u\partial_z) \bfP_u = (\Omega - K u)\bfe_3\times\bfP_u .
   \label{eq:eom-prec}
\end{align}
Integrating the above equation with weight $G(u)$ and using Eq.~\eqref{eq:cons} we obtain
\begin{align}
    \bfe_3\times(\Omega \bfN - K \bfF) = 0.
\end{align}
The above condition implies the following proportionality relation
\begin{align}
    \frac{F_1}{N_1} = \frac{F_2}{N_2} = \frac{\Omega}{K}.
    \label{eq:prop}
\end{align}
Meanwhile, subtracting Eq.~\eqref{eq:eom-prec} from Eq.~\eqref{eq:eom} we obtain
\begin{align}
    [\bfH_u - (\Omega-K u)\bfe_3]\times \bfP_u = 0,
\end{align}
which implies the alignment condition
\begin{align}
    \bfP_u = \frac{\tilde{\bfH}_u}{|\tilde{\bfH}_u|}\,\epsilon_u,
    \label{eq:alignment}
\end{align}
where $\epsilon_u=\pm1$ indicates whether $\bfP_u$ is aligned or anti-aligned with $\tilde{\bfH}_u=\bfH_u - (\Omega-K u)\bfe_3$. The wave solution is completely determined if $\bfP_u(t=0,z=0)$ and $(\Omega, K)$ are known. The proportionality relation in Eq.~\eqref{eq:prop} implies that $\bfe_3$, $\bfN$, and $\bfF$ are in the same plane. For simplicity, we choose the phase of the wave in Eq.~\eqref{eq:ansatz} to be $\phi=0$ so that
\begin{align}
    \bfN(t=0,z=0) &= N_\perp\bfe_1 + N_3\bfe_3 \\
    \bfF(t=0,z=0) &= F_\perp\bfe_1 + F_3\bfe_3, \\
    \intertext{and}
    \bfP_u(t=0,z=0) &=\tbP_u = \tP_{u,\perp} \bfe_1 + \tP_{u,3} \bfe_3,
\end{align}
where $N_\perp$, $N_3$, $F_\perp$, and $F_3$ are constants, and
\begin{widetext}
\begin{subequations}
    \label{eq:tP}
    \begin{align}
        \tP_{u,\perp} &= \frac{(N_\perp - F_\perp u)\epsilon_u}
        {\sqrt{[(N_3-\Omega) - (F_3-K)u]^2 + (N_\perp - F_\perp u)^2}}, \\
        \tP_{u,3} &= \frac{[(N_3-\Omega) - (F_3-K)u]\epsilon_u}
        {\sqrt{[(N_3-\Omega) - (F_3-K)u]^2 + (N_\perp - F_\perp u)^2}}.
        \label{eq:tP3}
    \end{align}
\end{subequations}
Integrating the above equations with weights $G(u)$ and $u G(u)$, respectively, we obtain the following self-consistent equations for the wave solution:
\begin{subequations}
    \label{eq:wave}
    \begin{align}
        N_\perp &= \int_{-1}^1 \frac{N_\perp - F_\perp u}
        {\sqrt{[(N_3-\Omega) - (F_3-K)u]^2 + (N_\perp - F_\perp u)^2}}\, G(u)\epsilon_u\,\rmd u, \label{eq:Np}\\
        F_\perp &= \int_{-1}^1 \frac{N_\perp - F_\perp u}
        {\sqrt{[(N_3-\Omega) - (F_3-K)u]^2 + (N_\perp - F_\perp u)^2}}\, u G(u)\epsilon_u\,\rmd u, \label{eq:Fp}\\
        N_3 &= \int_{-1}^1 \frac{(N_3-\Omega) - (F_3-K)u}
        {\sqrt{[(N_3-\Omega) - (F_3-K)u]^2 + (N_\perp - F_\perp u)^2}}\, G(u)\epsilon_u\,\rmd u, \label{eq:N3}\\
        F_3 &= \int_{-1}^1 \frac{(N_3-\Omega) - (F_3-K)u}
        {\sqrt{[(N_3-\Omega) - (F_3-K)u]^2 + (N_\perp - F_\perp u)^2}}\, u G(u)\epsilon_u\,\rmd u. \label{eq:F3}
    \end{align}
\end{subequations}
\end{widetext}
Eqs.~\eqref{eq:Np} and \eqref{eq:Fp} are not independent of each other for given $(\Omega, K)$ because of the proportionality relation in Eq.~\eqref{eq:prop}. One can solve one of these equations together with Eqs.~\eqref{eq:N3} and \eqref{eq:F3} for $\Omega$, $F_\perp$, and $F_3$ when $G(u)$, $K$, and $N_3$ are given.

\section{Numerical examples} \label{sec:examples}

The nonlinear equation set \eqref{eq:wave} is not guaranteed to have a nontrivial solution, although the solutions with the trivial configuration $\tP_{u,3}=1$ can be found by linearizing Eq.~\eqref{eq:eom} \cite{Izaguirre:2016gsx, Capozzi:2017gqd, Yi:2019hrp}. In this section, however, we demonstrate two cases where such wave solutions do exist. For simplicity, we will use a toy ELN distribution
\begin{align}
    G(u) = c_0 - c_1 u,
    \label{eq:ELN}
\end{align}
where $c_0$ and $c_1$ are constants. We will assume $c_0=0.5$ in both cases so that $N_3=1$ when $\tP_{u,3}=1$.

\subsection{ELN distribution without crossing}
\begin{figure}[htb]
    \begin{center}
        \includegraphics[trim=1 2 1 1, clip, scale=\figscale]{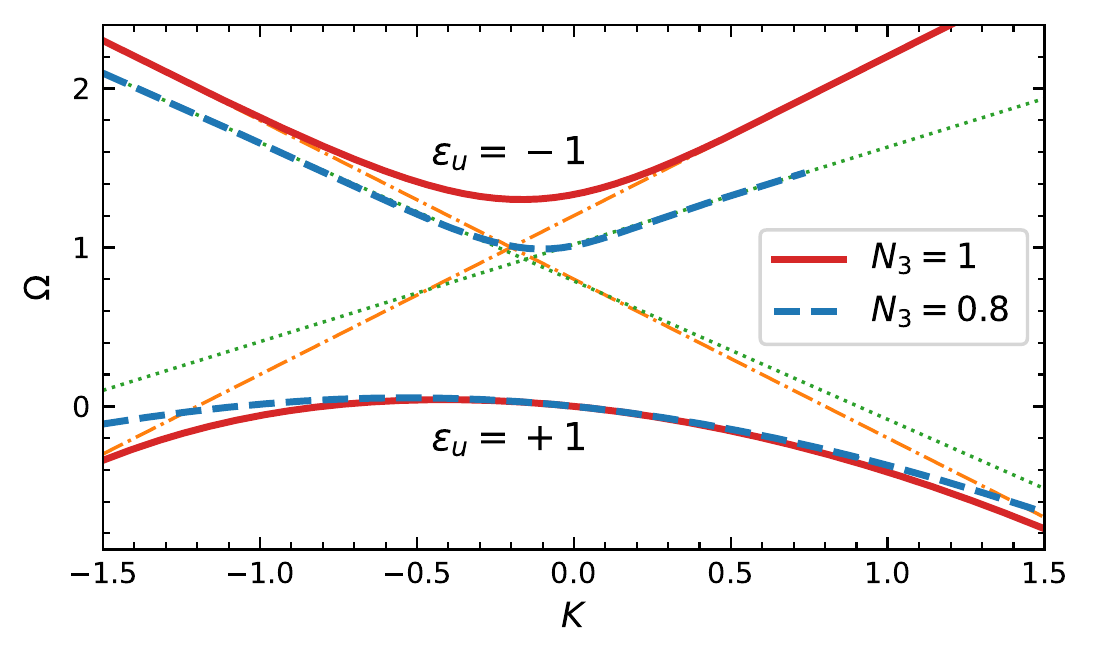}
    \end{center}
    \caption{The DRs of the flavor isospin waves in the 1D axisymmetric neutrino gas with the neutrino angular distribution $G(u)=0.5-0.3 u$ and the ELN density $N_3=1$ (solid line) and $N_3=0.8$ (dashed line), respectively. The upper and lower DR branches have the alignment signatures $\epsilon_u=-1$ and $+1$, respectively. The dot-dashed and dashed lines are the asymptotes of the DRs as the wave number $|K|\rightarrow\infty$.}
    \label{fig:DR-stable}
\end{figure}

In the first example, we consider the neutrino gas with $G(u) = 0.5 - 0.3 u$ which is always positive for $u\in[-1, 1]$.  We solve Eq.~\eqref{eq:wave} for the wave solution with $N_3=0.8$ and $\epsilon_u=\pm1$ and show its DR in Fig.~\ref{fig:DR-stable}. 

\begin{figure*}
    \begin{center}
        $\begin{array}{@{}c@{\hspace{0.1in}}c@{}}
            \includegraphics[trim=1 1 1 1, clip, scale=\figscale]{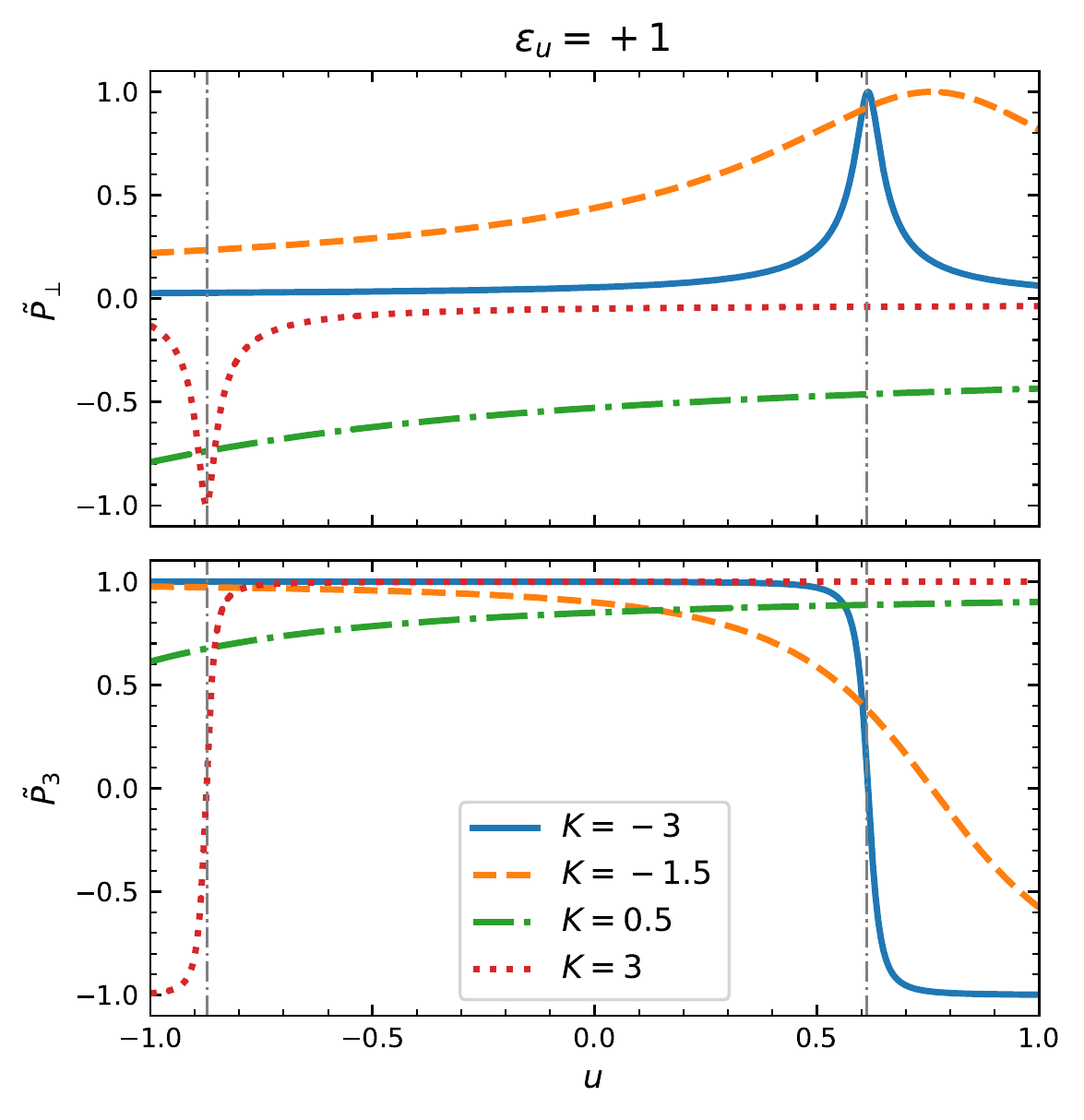} &
            \includegraphics[trim=20 1 1 1, clip, scale=\figscale]{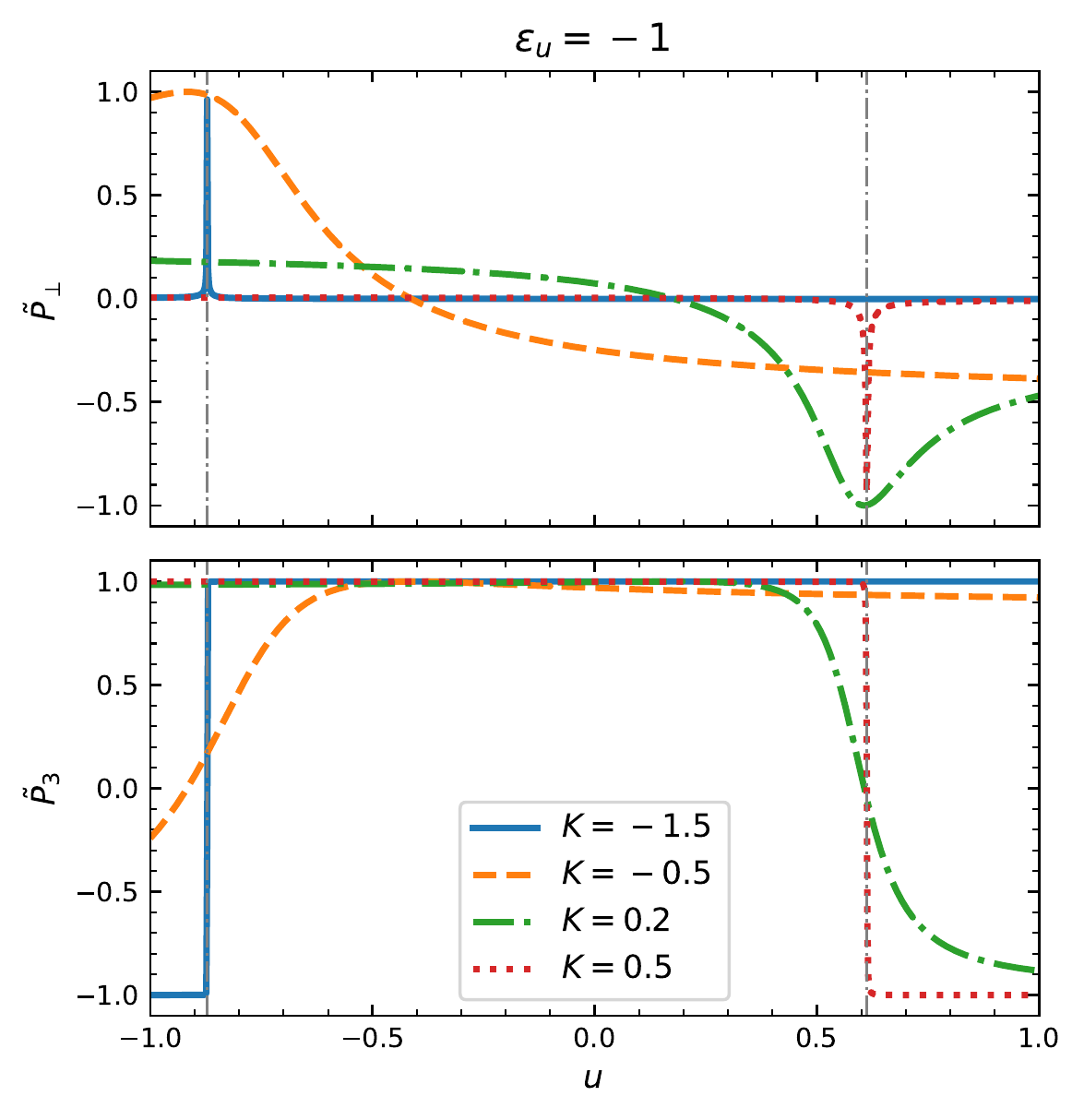} 
            \end{array}$
    \end{center}
    \caption{The flavor Bloch vectors of a few nontrivial wave solutions (with the wave numbers $K$ as labeled) as functions of the neutrino velocity $u$ (along the $z$ axis). The upper and lower panels show the components of the Bloch vectors that are perpendicular and parallel to the flavor axis, respectively. The solutions in left and right panels are on the DR branches shown in Fig.~\ref{fig:DR-stable} with the alignment signatures $\epsilon_u=+1$ and $-1$, respectively. The thin vertical dot dashed lines indicate the locations of the angular spectral split predicted by Eq.~\eqref{eq:split}.}
    \label{fig:P-stable}
\end{figure*}

Obviously, the wave solution with $N_3=0.8$ cannot have $\tP_{u,3}=1$ for all $u$'s. In Fig.~\ref{fig:P-stable} we plot $\tP_{u,\perp}$ and $\tP_{u,3}$ of this solution that are determined by Eq.~\eqref{eq:tP}. As one can see, $\tP_{u,3}$ approaches a step function as $|K|$ and $|\Omega|$ become large. This behavior is actually expected from Eqs.~\eqref{eq:tP} and \eqref{eq:wave}. As $|\Omega|$ and $|K|$ approach infinity, both $N_\perp$ and $F_\perp$ are small, which in turn implies that $\tP_{u,\perp}\rightarrow 0$ and 
\begin{align}
    \tP_{u,3} \longrightarrow \mathrm{sgn}[(K-K_0)(u-\us)\epsilon_u],
    \label{eq:tP3-lim}
\end{align}
where 
\begin{subequations}
    \begin{align}
        \Omega_0 &= \lim_{|K|\rightarrow\infty}\int_{-1}^1 G(u)\tP_{u,3}\,\rmd u, \\
        \intertext{and}
        K_0 &= \lim_{|K|\rightarrow\infty}\int_{-1}^1 u G(u)\tP_{u,3}\,\rmd u.
    \end{align}        
\end{subequations}
are the asymptotic limits of the wave frequency and wave number, respectively, and
\begin{align}
    \us=\frac{\Omega-\Omega_0}{K-K_0}
    \label{eq:us}
\end{align}
is the value of $u$ where $\tP_{u,3}$ transitions between $+1$ and $-1$. We note that $\Omega_0=N_3$ is one of the parameters that we used to compute the wave solutions. As $\tP_{u,\perp}\rightarrow 0$, it is numerically difficult to solve Eq.~\eqref{eq:wave}. For this reason, we are able to plot only part of the DR of the wave solution with $N_3=0.8$ and $\epsilon_u=-1$ in Fig.~\ref{fig:DR-stable}.

The step-like structure of $\tP_{u,3}$ at $|K|\rightarrow1$ is similar to the swap/split phenomenon of the neutrino energy spectra \cite{Duan:2006an,Raffelt:2007cb} except here it occurs in the angular dimension. The location of the angular spectral split $\us$ can be determined by the asymptotic behavior of $\tP_{u,3}$ in Eq.~\eqref{eq:tP3-lim} and the ELN density $N_3$:
\begin{align}
    N_3 = \pm\left[  \int_{\us}^1 G(u)\epsilon_u\,\rmd u - \int_{-1}^{\us} G(u)\epsilon_u\,\rmd u \right].
\end{align}
For the simple angular distribution in Eq.~\eqref{eq:ELN} and constant $\epsilon_u$ for all neutrinos, one finds two possible splitting points
\begin{align}
    \us_\pm = \pfrac{c_0}{c_1} - \sqrt{\pfrac{c_0}{c_1}^2 +1 \mp\pfrac{N_3}{c_1}}.
    \label{eq:split}
\end{align}
We plot $u=\us_\pm$ as the vertical dot-dashed lines in Fig.~\ref{fig:P-stable} which indeed pinpoint the spectral split locations at large $|K|$. 

Eq.~\eqref{eq:split} gives two values of $K_0$ at $\Omega=N_3$:
\begin{align}
    K_{0,\pm} = \pm c_0[(\us_\pm)^2-1] \mp \frac{2 c_1}{3} (\us_\pm)^3.
    \label{eq:K0}
\end{align} 
Using thee values, we plot Eq.~\eqref{eq:us} for $N_3=0.8$ as the dotted lines in Fig.~\ref{fig:DR-stable} which are the asymptotes of the DR branches of the nontrivial wave solutions.

In Eqs.~\eqref{eq:split} and \eqref{eq:K0}, 
\begin{subequations}
    \begin{align}
        \us_\pm &\longrightarrow \pm1,\\
        \intertext{and}
        K_{0,\pm}&\longrightarrow\int_{-1}^1 G(u)u\,\rmd u
    \end{align}
\end{subequations}
in the limit $N_3\rightarrow c_0$. In this limit, the wave solution has the trivial configuration $\tP_{u,3}=1$. It is numerically difficult to obtain the DR of a trivial wave solution from Eq.~\eqref{eq:wave} directly. Instead, we obtained the DR branches of the trivial wave solutions by linearizing Eq.~\eqref{eq:eom} \cite{Izaguirre:2016gsx}. These DR branches together with their asymptotes are shown in Fig.~\ref{fig:DR-stable}. 

\begin{figure*}
    \begin{center}
        $\begin{array}{@{}c@{\hspace{0.1in}}c@{}}
            \includegraphics[trim=1 1 50 1, clip, scale=\figscale]{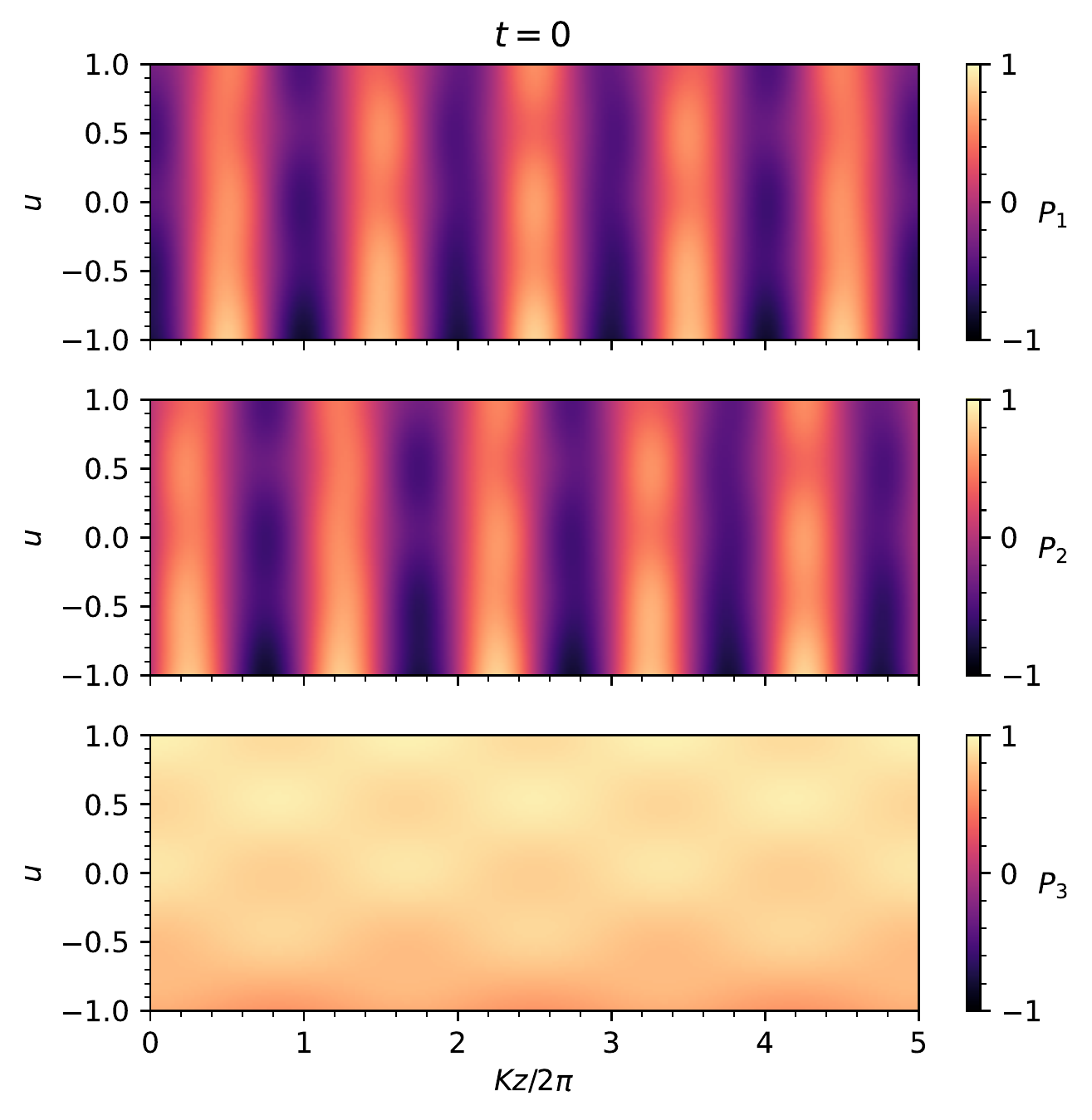} &
            \includegraphics[trim=25 1 1 1, clip, scale=\figscale]{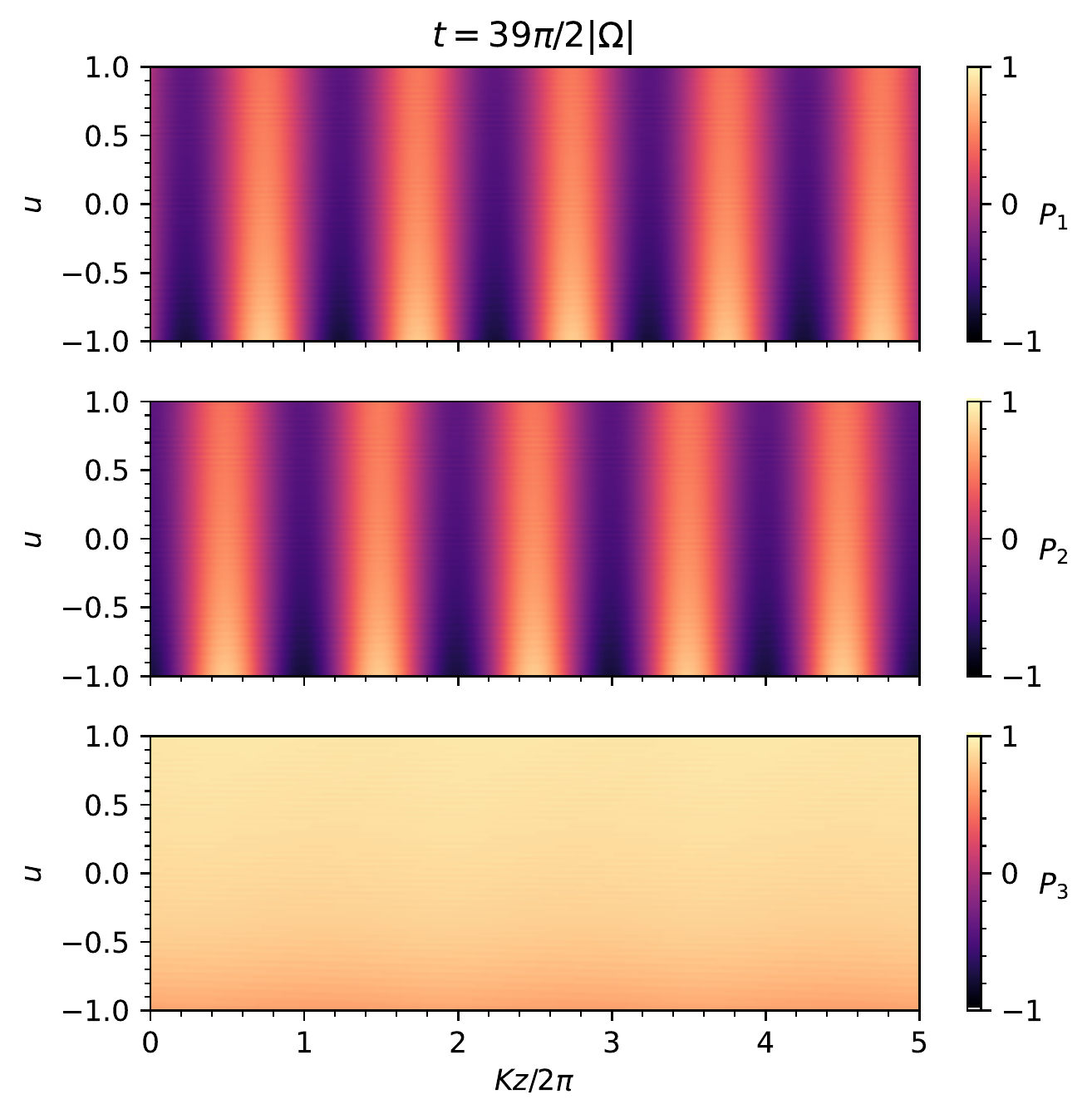}  
            \end{array}$
    \end{center}
    \caption{The three components of the flavor Bloch vector $\bfP_u(t,z)$ (shown as the colors of the pixels in the upper, middle, and lower panels) as functions of the neutrino velocity $u$ and coordinate $z$ at time $t=0$ (left panels) and $39\pi/2|\Omega|$ (right panels) in a perturbed wave calculation, respectively. At $t=0$, the neutrino gas is slightly perturbed from the pure wave configuration which has wave number $K=0.5$, frequency $\Omega\approx-0.144$, and alignment signature $\epsilon_u=+1$ (dot-dashed lines in the left panels of Fig.~\ref{fig:P-stable}).}
    \label{fig:wave-ptrb}
\end{figure*}

To demonstrate that the solution to Eqs.~\eqref{eq:tP} and \eqref{eq:wave} indeed describes a propagating wave in the neutrino gas, we performed two calculations in a 1D periodic box of length $L=10\pi/|K|$ and with the initial conditions
\begin{subequations}
    \label{eq:IC-stable}
    \begin{align}
        P_{u,1}(0, z) &= P_{u,\perp}(z) \cos(K z), \\
        P_{u,2}(0, z) &= -P_{u,\perp}(z) \sin(K z), \\
        P_{u,3}(0, z) &= \tP_{u,3} + \delta a \cos(2\pi u) \cos(3 K z/5),
    \end{align}        
\end{subequations}
where $\tP_{u,3}$ is solved from Eqs.~\eqref{eq:tP} and \eqref{eq:wave} with $N_3=0.8$, $\epsilon_u=+1$, and $K=0.5$,%
\footnote{This nontrivial wave solution corresponds to the dot-dashed lines in left panels of Fig.~\ref{fig:P-stable} which has $\Omega\approx-0.144$, $F_\perp\approx 0.169$, and $F_3\approx-0.121$. For this solution, $\tP_{u,\perp}=\pm\sqrt{1-\tP_{u,3}^2}$ where the sign depends on the choice of $\bfe_1$.}
and $P_{u,\perp}(z)=-\sqrt{1-P_{u,3}^2(0, z)}$. We take $\delta a=0$ and $0.05$ in the two calculations which produce the exact and perturbed wave solutions, respectively. We have chosen the form of the perturbation such that $N_3=0.8$ for all $z$. Both calculations were carried out using the \nugas\ Python package \cite{nugas} with 1000 equally spaced $z$ bins and 1001 equally spaced $u$ bins. The integral over $u$ is performed using the composite Simpson's rule \cite{NR}.

In the calculation with $\delta a=0$, $\bfP_u(t,z)$ simply translates along $z$ with the phase velocity $V=\Omega/K$ as time progresses. In the perturbed wave calculation, the general wave structure is still preserved and propagates in space with time. We show the initial and a later snapshots of the flavor Bloch vectors of the neutrino gas in the perturbed wave calculation in Fig.~\ref{fig:wave-ptrb}.

\begin{figure*}
    \begin{center}
        $\begin{array}{@{}c@{\hspace{0.1in}}c@{}}
            \includegraphics[trim=1 1 1 1, clip, scale=\figscale]{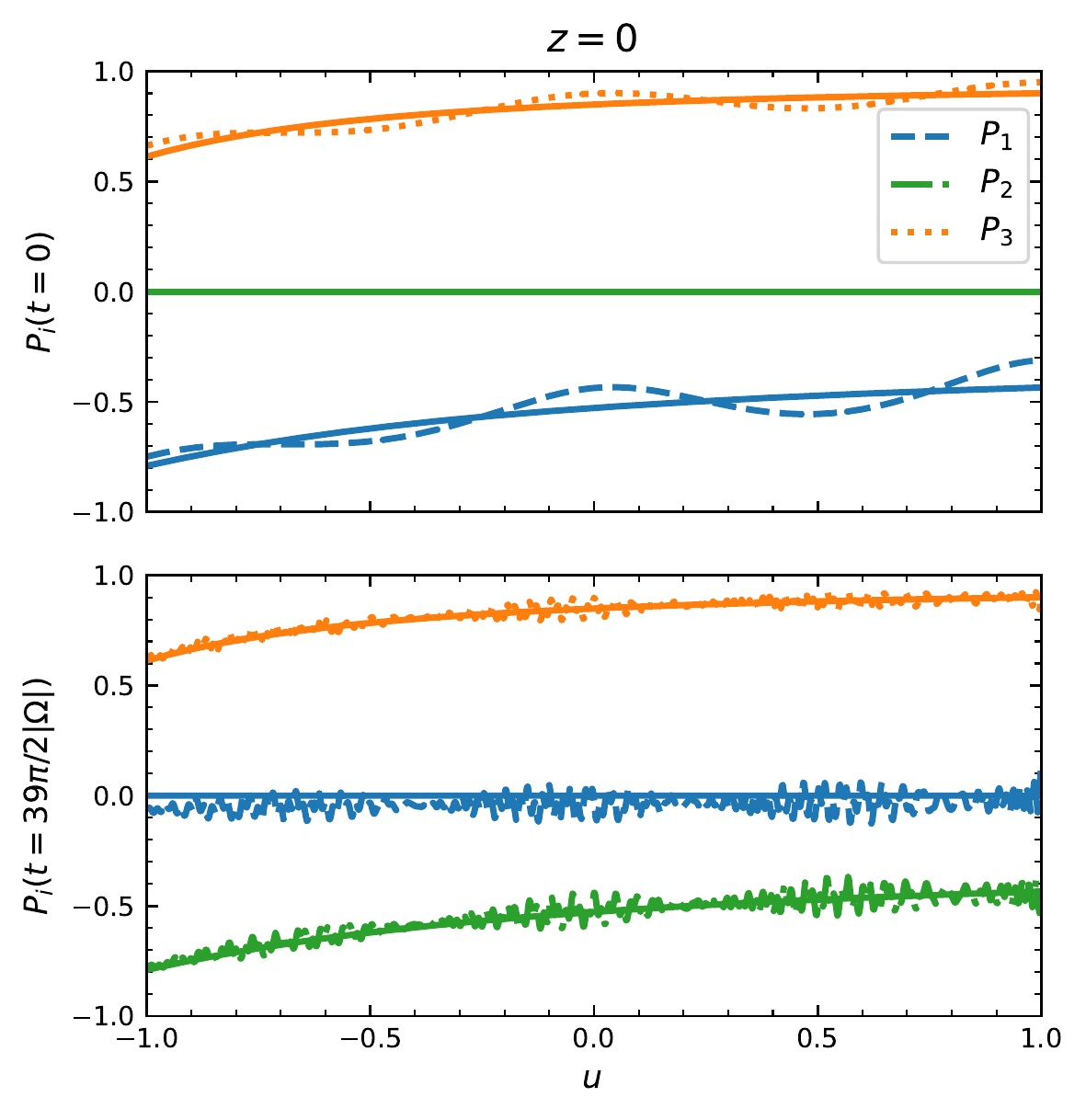} &
            \includegraphics[trim=20 1 1 1, clip, scale=\figscale]{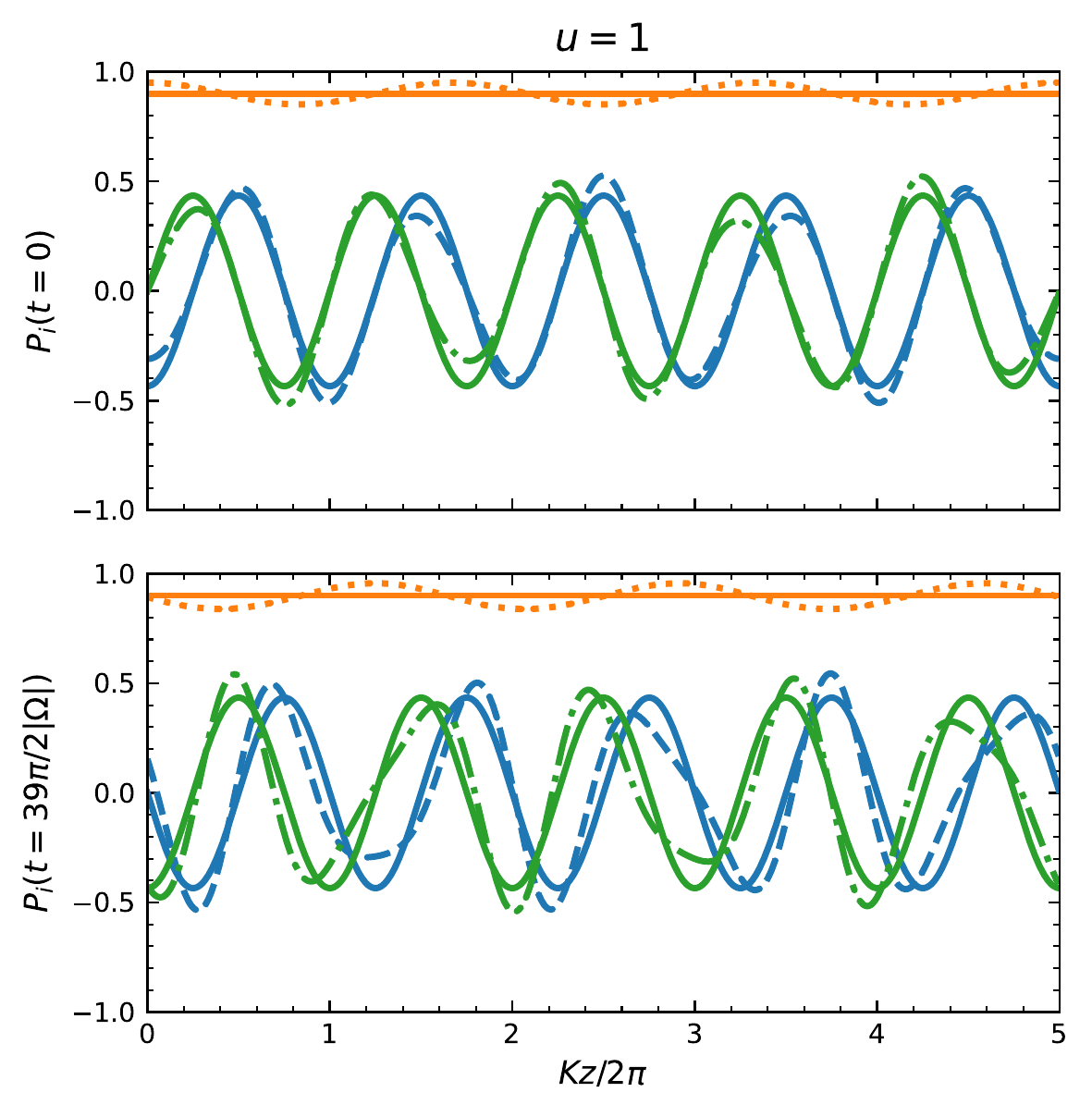} 
            \end{array}$
    \end{center}
    \caption{The three components of the flavor polarization vector $\bfP_u$ (as labeled) as functions of neutrino velocity $u$ at coordinate $z=0$ (left panels) and as functions of $z$ at $u=1$ (right panels) in the perturbed wave calculation shown in Fig.~\ref{fig:wave-ptrb} at $t=0$ (upper panels) and $39\pi/2|\Omega|$ (lower panels), respectively. The solid curves are the corresponding quantities in the exact wave calculation.}
    \label{fig:P-stable-ptrb}
\end{figure*}

In Fig.~\ref{fig:P-stable-ptrb} we compare the polarization vectors in the exact and perturbed wave calculations in the angular dimension at $z=0$ (left panels) and also in the spatial dimension at $u=1$ (right panels), respectively. The perturbations in the perturbed wave calculation do not grow substantially up to $t=39\pi/2|\Omega|$, although small angular structures develop over time. It is interesting to note that the perturbations remain smooth in the spatial dimension.

\subsection{ELN distribution with crossing}

\begin{figure}[htb]
    \begin{center}
        \includegraphics[trim=1 2 1 1, clip, scale=\figscale]{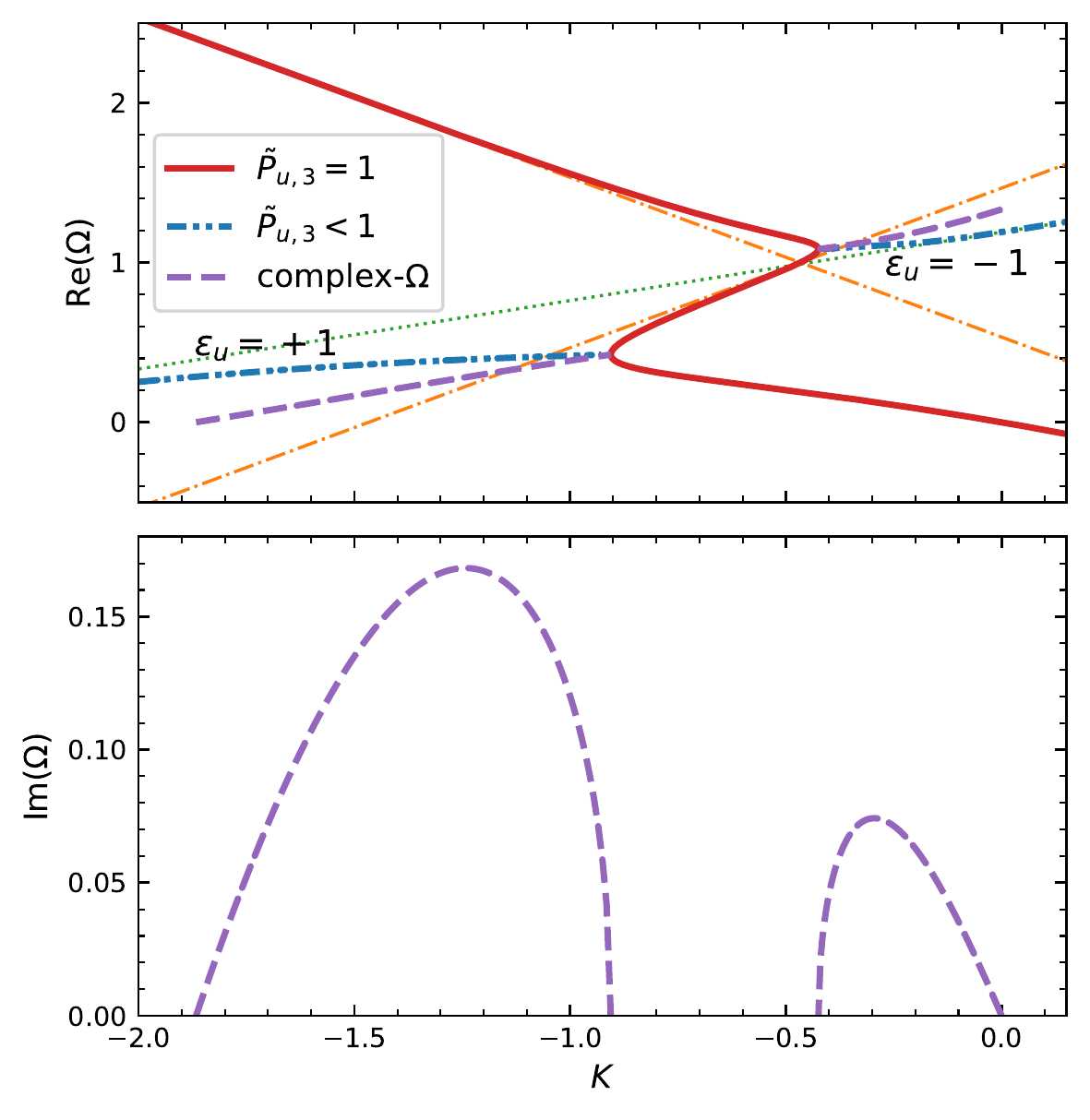}
    \end{center}
    \caption{The various DRs branches of the flavor isospin wave in the 1D axisymmetric neutrino gas with the ELN distribution $G(u)=0.5-0.7u$ and ELN density $N_3=1$. The solid and dashed lines are the DR branches of the trivial wave solutions with the real and complex wave frequencies $\Omega$, respectively. The dot-dot-dashed lines are the DR branches of the nontrivial wave solutions. The thin dot-dashed and dotted lines are the asymptotes of the DR branches with real frequencies. See the text for details.}
    \label{fig:DR-unstable}
\end{figure}

As shown in Ref.~\cite{Yi:2019hrp}, the two DR branches of the trivial flavor isospin wave with $\tP_{u,3}=1$ merge into one when the ELN distribution has a (single) crossing%
\footnote{Because the two DR branches have different alignment signatures $\epsilon_u=\pm1$, they must merge at the point where $\tilde{\bfH}=(\bfN-u\bfF)-(\Omega-u K)\bfe_3=\mathbf{0}$ for all $u$ which gives $\Omega=\Omega_0=N_3$ and $K=K_0=F_3$ at this point.}. 
At the same time, the trivial wave solutions with complex frequencies $\Omega$ appear. The flavor isospin wave with a positive $\mathrm{Im}(\Omega)$ is unstable and its flavor Bloch vectors $\bfP_u$ deviate from $\bfe_3$ exponentially as time progresses. The complex-$\Omega$ DR branches of the trivial wave solutions sprout from the turning point of their real DR branches and extend to the points that are determined by the crossing point of the ELN distribution. As an example, we computed these DR branches for the ELN distribution $G(u) = 0.5 - 0.7 u$ and plot them in Fig.~\ref{fig:DR-unstable}.%
\footnote{We did not plot the DR branches of the trivial wave solutions with complex wave numbers because they are not relevant for the discussion here.}

Because $N_3 = 1 < \int_{-1}^1 |G(u)|\,\rmd u$, it is possible to have an angular split at $\us_+$ so that
\begin{align}
    \int_{-1}^{\us_+} G(u)\,\rmd u - \int_{\us_+}^1 G(u)\,\rmd u = N_3.
    \label{eq:us2}
\end{align}
This suggests that there exist nontrivial wave solutions that have $N_3=1$ and yet $\tP_{u,3}<1$. Indeed, solving Eq.~\eqref{eq:wave} we found two real DR branches of the nontrivial wave solutions and plot them in Fig.~\ref{fig:DR-unstable}. The new DR branches have opposite alignment signatures and stem from the same turning points of the real DR  branch of the trivial wave solution where the complex-$\Omega$ DR branches start, and they extend toward the asymptotes $\Omega=\us_+(K-K_{0,+}) + \Omega_0$, where $\Omega_0=N_3$ and $K_{0,+}$ is given by Eq.~\eqref{eq:K0}.

\begin{figure*}
    \begin{center}
        $\begin{array}{@{}c@{\hspace{0.1in}}c@{}}
            \includegraphics[trim=1 1 1 1, clip, scale=\figscale]{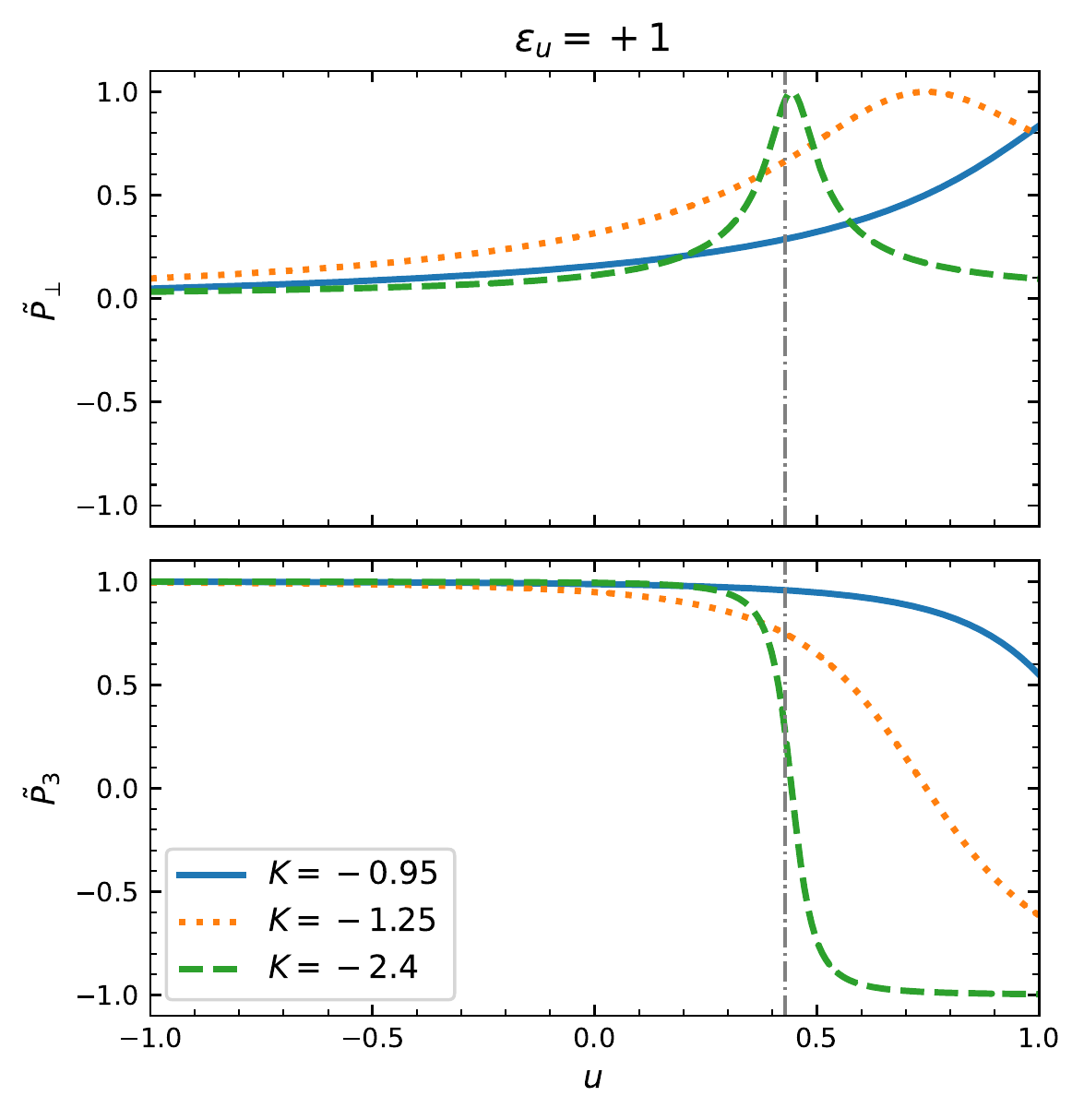} &
            \includegraphics[trim=20 1 1 1, clip, scale=\figscale]{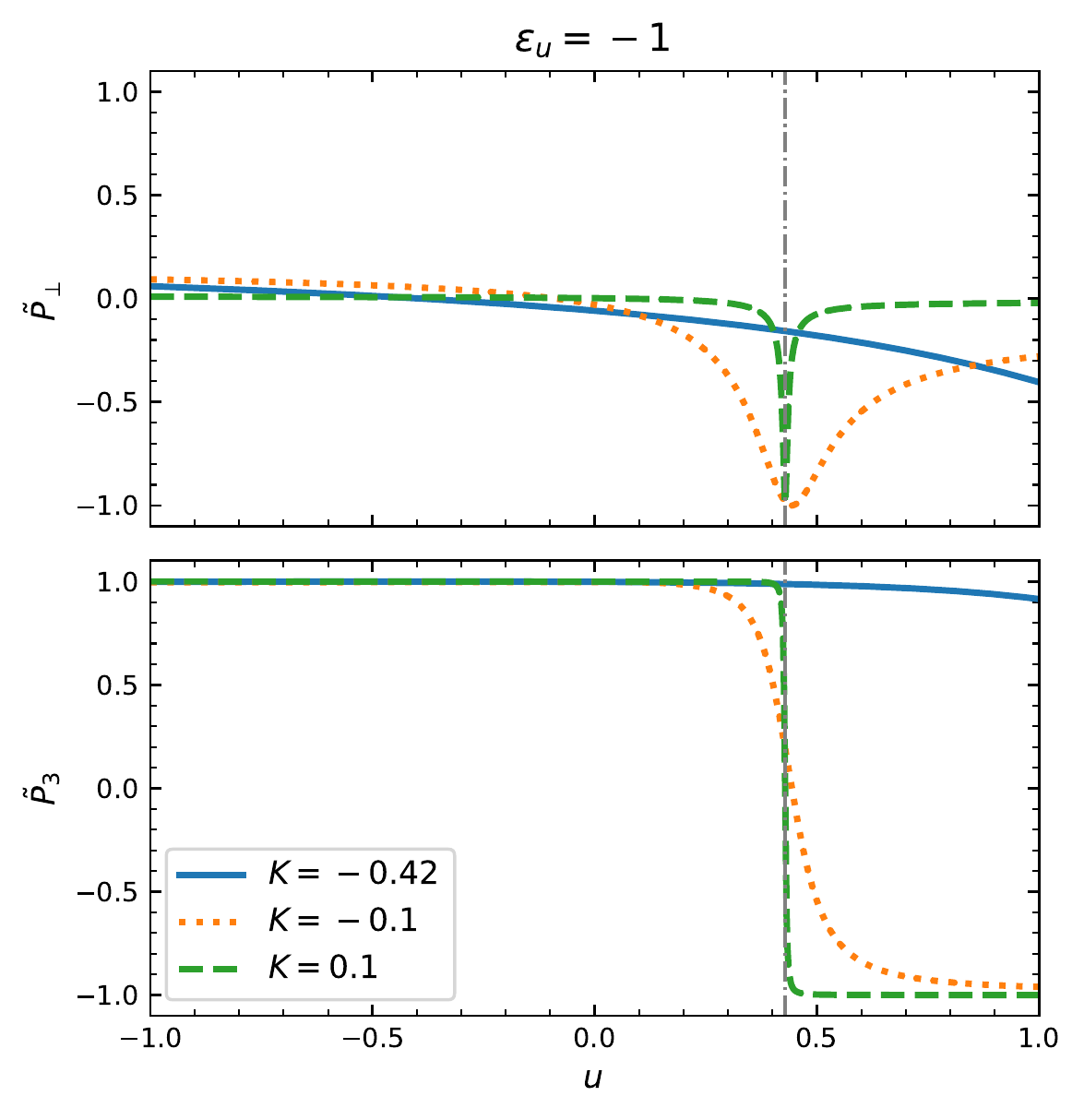} 
            \end{array}$
    \end{center}
    \caption{Similar to Fig.~\ref{fig:P-stable} but for the nontrivial wave solutions shown in Fig.~\ref{fig:DR-unstable}. The thin vertical dot dashed lines indicate the location of the angular spectral split $\us_+$ predicted by Eq.~\eqref{eq:us2}.}
    \label{fig:P-unstable}
\end{figure*}

In Fig.~\ref{fig:P-unstable} we show $\tP_{u,\perp}$ and $\tP_{u,3}$ of the nontrivial wave solutions with a few wave numbers. When the DR branches of the nontrivial solution approach the nexus point with those of the trivial solution, $\tP_{u,3}\rightarrow 1$ and $\tP_{u,\perp}\rightarrow 0$ as expected. As they extend away from the nexus point, $\tP_{u,3}$ again approaches a step-like function of $u$, and the splitting point $\us_+$ is determined by Eq.~\eqref{eq:us2}.

To demonstrate the dynamic evolution of the neutrino gas with an ELN crossing, we performed two calculations with the aforementioned ELN distribution in a periodic box of length $L=10\pi/|K|$ and with the following initial conditions:
\begin{subequations}
    \label{eq:IC-unstable}
    \begin{align}
        P_{u,1}(0, z) &= P_{u,\perp}(z) \cos(K z), \\
        P_{u,2}(0, z) &= -P_{u,\perp}(z) \sin(K z), \\
        P_{u,3}(0, z) &= (1-\delta a) + \delta a \tP_{u,3},
    \end{align}
\end{subequations}
where $\tP_{u,3}<1$ is solved from Eqs.~\eqref{eq:tP} and \eqref{eq:wave} with $K=-1.25$ and $\epsilon=+1$,%
\footnote{This wave solution corresponds to the dotted curves in the left panels of Fig.~\ref{fig:P-unstable} which has $\Omega\approx0.392$, $F_\perp\approx0.064$, and $F_3\approx-0.435$.} 
and $P_{u,\perp}(z)=-\sqrt{1-P_{u,3}^2(0,z)}$. We take $\delta a=10^{-2}$ and $10^{-4}$, respectively, in the two calculations. We note that the initial conditions in Eq.~\eqref{eq:IC-unstable} mix the trivial and nontrivial wave solutions both with the ELN density $N_3=1$. 

\begin{figure}
    \begin{center}
        \includegraphics[trim=1 1 1 1, clip, scale=\figscale]{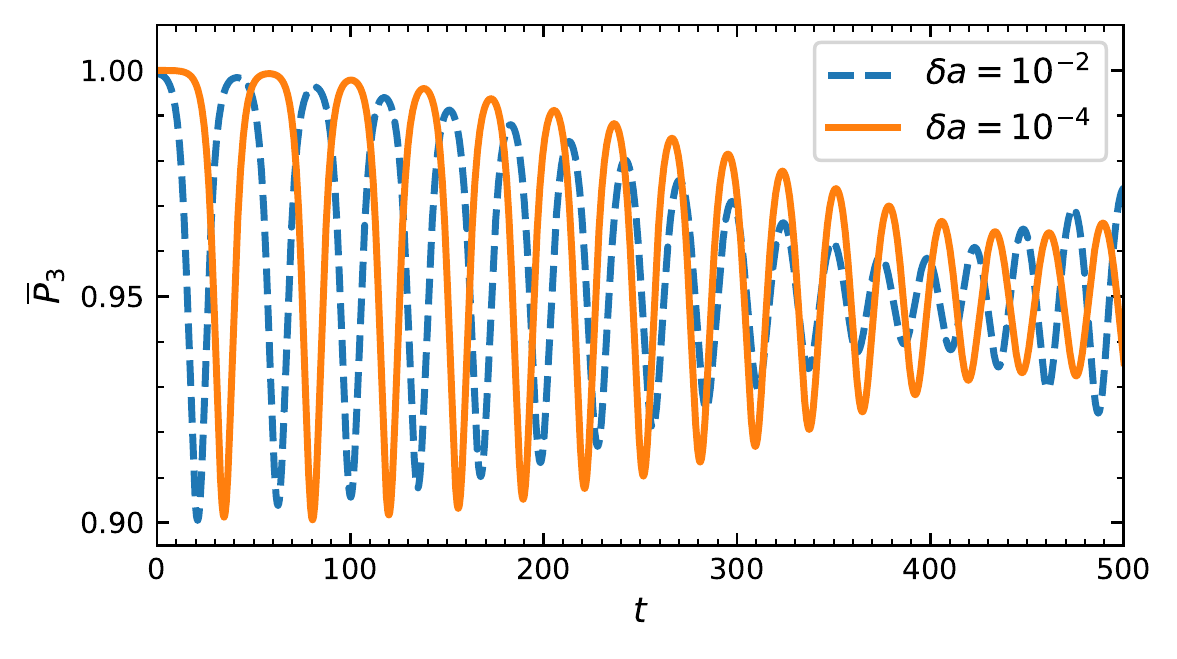}
    \end{center}
    \caption{The angular and spatial average of the third components of the flavor Bloch vectors of two neutrino gases as functions of time $t$. At $t=0$ both gases are in the weak-interaction states mixed with a nontrivial wave solution of wave number $K=-1.25$ and alignment signature $\epsilon_u=+1$ (dotted curves in the left panels of Fig.~\ref{fig:P-unstable}). The initial amplitudes of the wave component are $\delta a=10^{-2}$ and $10^{-4}$ for the two gases. See the text for details.}
    \label{fig:pendulum}
\end{figure}

We define
\begin{align}
    \overline{P}_3\equiv \frac{\int_0^L\rmd z\int_{-1}^1\rmd u P_{u,3}(t,z)|G(u)|}{L\int_{-1}^1 |G(u)|\,\rmd u}
    \label{eq:avgP3}
\end{align}
as a measure of the average deviation of $\bfP_u$ from $\bfe_3$ at time $t$. We plot $\avgP_3(t)$ of the two calculations in Fig.~\ref{fig:pendulum}. In both calculations,  $\overline{P}$ oscillate semi-periodically with time, but the neutrino gas with a smaller mixing of the nontrivial wave solution at $t=0$ has a longer oscillation period. The oscillation amplitudes in both cases are first damped with time but then grow back.

\begin{figure*}
    \begin{center}
        $\begin{array}{@{}c@{\hspace{0.1in}}c@{}}
            \includegraphics[trim=1 1 50 1, clip, scale=\figscale]{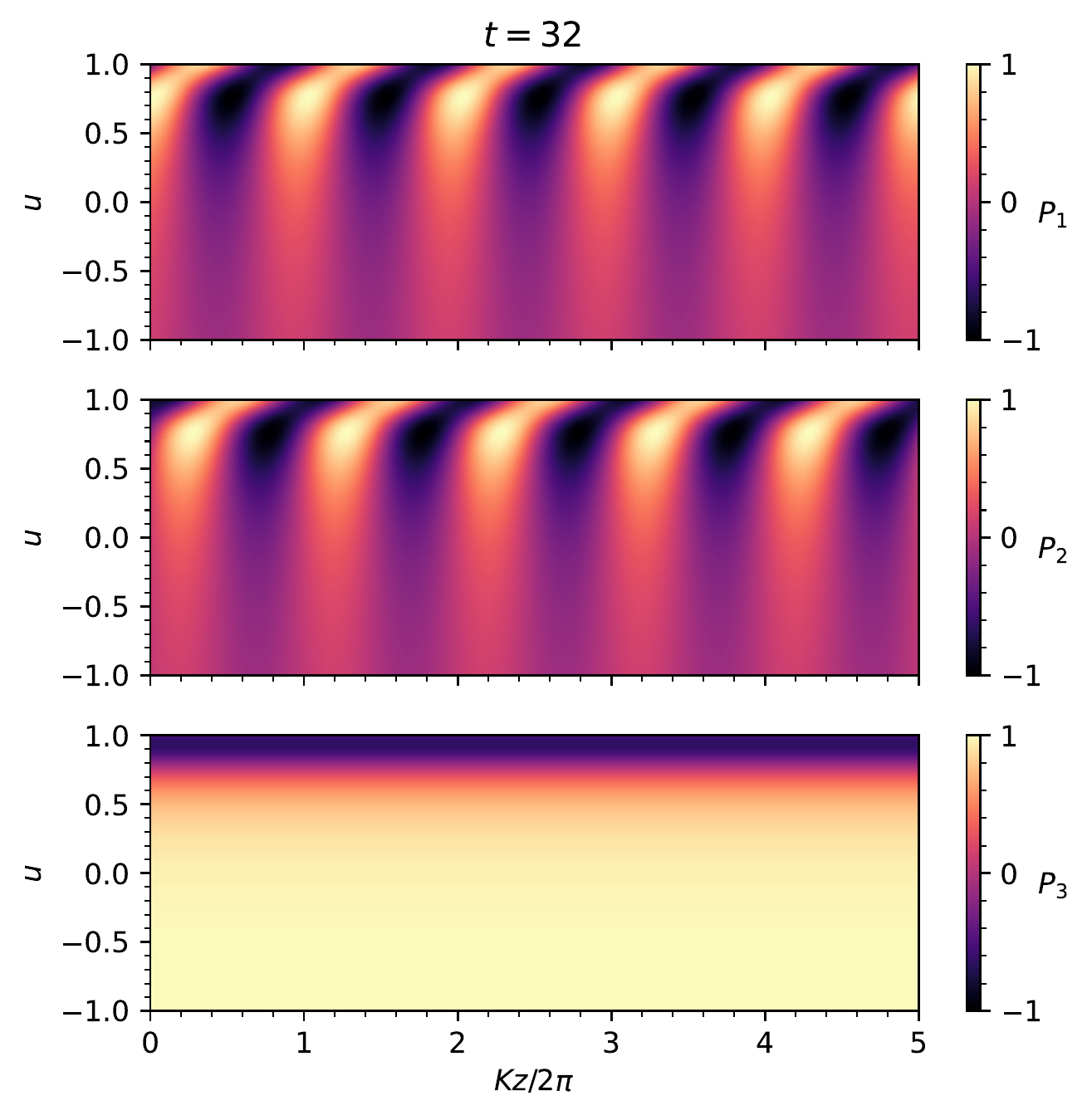} &
            \includegraphics[trim=25 1 1 1, clip, scale=\figscale]{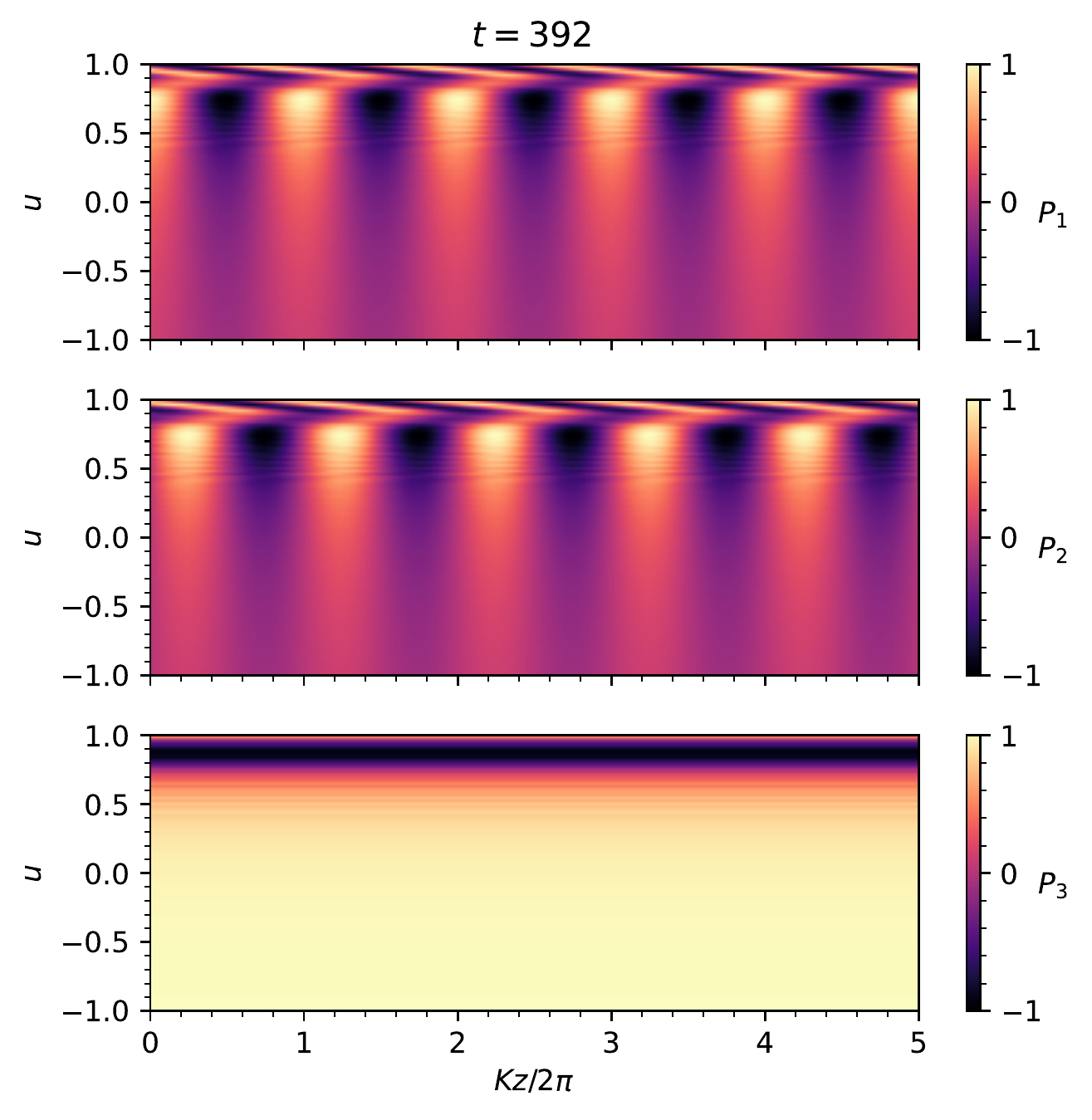}  
            \end{array}$
    \end{center}
    \caption{The three components of the flavor Bloch vector $\bfP_u(t,z)$ at $t=32$ (left panels) and $392$ (right panels) for the neutrino gas that has a small mixing of a nontrivial flavor isospin wave with amplitude $\delta a=10^{-4}$ initially (solid curve in Fig.~\ref{fig:pendulum}).}
    \label{fig:wave-ptrb-unstable}
\end{figure*}

\begin{figure*}
    \begin{center}
        \includegraphics[trim=1 1 1 1, clip, scale=\figscale]{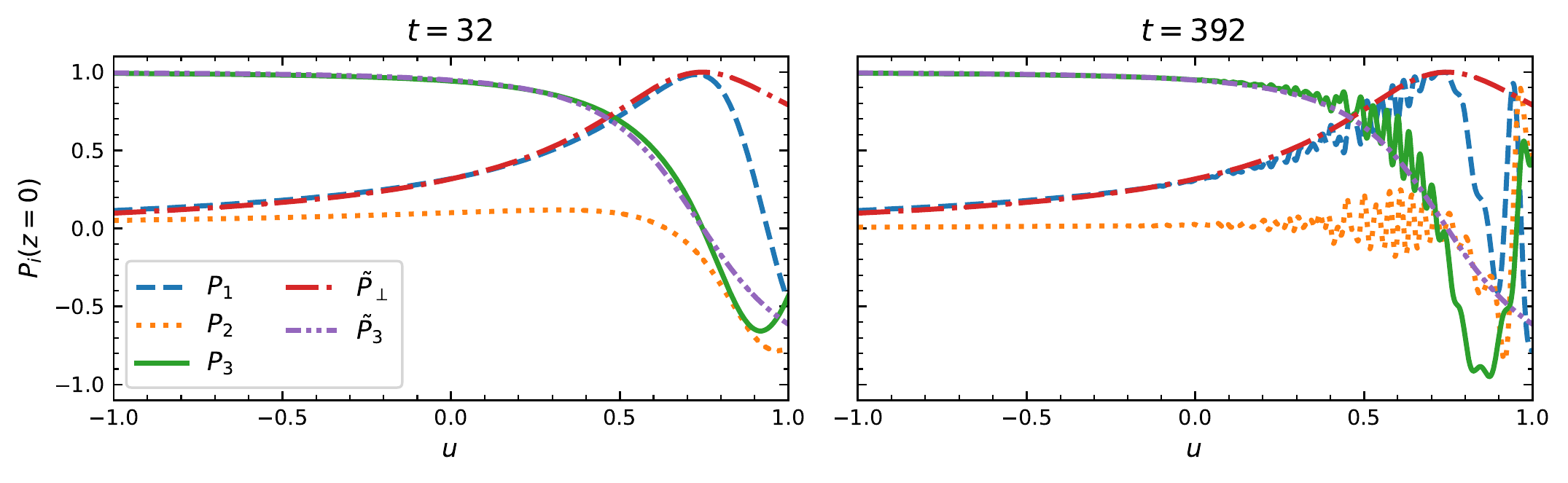}
    \end{center}
    \caption{Same as Fig.~\ref{fig:wave-ptrb-unstable} but for the flavor Bloch vectors at $z=0$ only. The dot-dashed and dot-dot-dashed lines represent the nontrivial wave solution which is solved from Eqs.~\eqref{eq:tP} and \eqref{eq:wave}.}
    \label{fig:P-unstable-ptrb}
\end{figure*}

In Figs.~\ref{fig:wave-ptrb-unstable} and \ref{fig:P-unstable-ptrb} we show the snapshots of $\bfP_u$ in the calculation with a smaller mixing ($\delta a=10^{-4}$) near the first local minimum of $\avgP_3(t)$ and another minimum at a later time, respectively. Near these minima, the neutrino gas clearly develops a wave pattern similar to the flavor isospin wave solution solved from Eqs.~\eqref{eq:tP} and \eqref{eq:wave}. This suggests that the semi-periodic behavior of the neutrino gas shown in Fig.~\ref{fig:pendulum} describes an oscillation of the neutrino gas between its initial condition $P_{u,3}\approx1$ and the nontrivial wave solution with $\tP_{u,3}<1$. 
 
\section{Discussion and conclusions}\label{sec:conclusions}

We have shown that there indeed exist nontrivial flavor isospin waves in dense 1D axisymmetric neutrino gases as was predicted in Ref.~\cite{Duan:2008fd}. For a monochromatic wave, the flavor Bloch vectors $\bfP_u$ of all neutrinos with the velocity $u$ (along the $z$ axis) precess uniformly in space and time in a way similar to a magnetic spin wave in a lattice of magnetic dipoles. The DRs for fast neutrino flavor conversions obtained earlier \cite{Izaguirre:2016gsx} correspond to those of the flavor isospin waves with a trivial configuration $\bfP_u=\bfe_3$. We have derived a set of self-consistent equations from which one can compute the characteristic quantities such as the frequency and wave number of a monochromatic, nontrivial flavor isospin wave. Using these equations we have computed the DRs of the flavor isospin waves in two representative cases which have ELN distributions $G(u)$ with and without a crossing, respectively.

Generally speaking, there is no fast flavor isospin wave solution with a nontrivial configuration, i.e. $|P_{u,3}|<1$, if all the neutrinos begin in (approximately) weak-interaction states and the ELN distribution has no crossing. If there is a shallow crossing in the ELN distribution as in the second example that we have shown in Sec.~\ref{sec:examples}, there exist nontrivial flavor isospin waves whose DR branches join the real DR branches of the trivial wave solutions at the points where their complex-frequency DR branches sprout. The flavor Bloch vectors of the nontrivial wave solutions form angular spectral split when the wave number is large.

We have demonstrated the nontrivial flavor isospin waves in a few numerical examples. In the example with a positive neutrino angular distribution $G(u)$ and an ELN density $N_3<\int_{-1}^1 G(u)\rmd u$, we showed that the nontrivial flavor isospin wave seems to be stable when it is perturbed slightly. In another example with a crossed ELN distribution, the neutrino gas seems to oscillate between a nontrivial wave solution and its initial state where the neutrinos are all (approximately) in the weak-interaction states.

The existence of nontrivial flavor isospin wave is important because it reveals a possible and interesting outcome of fast neutrino oscillations in the nonlinear regime. Such outcome has been suggested in some of the calculations \cite{Martin:2019gxb,Martin:2021xyl,Richers:2021xtf}. However, many open questions remain to be answered. For example, we have only showcased the nontrivial flavor isospin wave solutions whose DRs are closely related to those of the trivial wave solutions. Because of the nonlinear nature of the EoM, it is possible that there exist other nontrivial wave solutions with very different DRs, e.g.\ with different Bloch vector alignment signatures on the same DR branch but at different wave numbers. Another open question is the existence of nontrivial flavor isospin waves in a neutrino gas whose ELN distribution has a very deep crossing. It has been shown in Ref.~\cite{Yi:2019hrp} that the complex-$\Omega$ DR branches of the trivial wave solution will merge when the ELN crossing is deep enough, and the corresponding real DR branch can  disappear when the crossing gets even deeper. Although we have found some nontrivial wave solutions in such scenarios, we do not have a good insight of their behaviors at the moment.

Because of their nonlinear nature, it is not obvious how the flavor isospin waves of different wave numbers may interfere with each other. We have performed very few numerical calculations whose initial configurations are perturbed in a way without changing the ELN density. These calculations, of course, do not prove whether such waves are stable against arbitrary perturbations. In all our calculations, the flavor configuration of the neutrino gases do develop smaller and smaller angular structures as time progresses while the spatial perturbations remain smooth. Such fine flavor structures can be difficult to resolve and pose a challenge to accurate, long-term numerical calculations. Although it is possible that the coherent wave pattern may be destroyed in some cases as suggested in other neutrino gas models \cite{Raffelt:2007yz,Johns:2020qsk} and the flavor equilibrium is resulted, it is by no means a sure conclusion for all neutrino gases. 

\begin{acknowledgments}
We thank S.~Richers for the useful discussion. 
This material is based upon work supported by the U.S.\ Department of Energy, Office of Science, Office of Nuclear Physics under Award Number DE-SC0017803 (H.D.\ and S.O.) and Contract Number DE-AC52-06NA25396 (J.D.M.).
\end{acknowledgments}

\appendix
\section{NuGas Python Package and the Lax42 Algorithm}
\newcommand{\tf}{\tilde{f}}
\newcommand{\tg}{\tilde{g}}
\newcommand{\tfz}{\widetilde{[\partial_z f]}}
We have developed a Python package \nugas\ to compute collective flavor oscillations in a few neutrino gas models and made it publicly available \cite{nugas}. In particular, the subpackage \texttt{f2e0d1a} solves Eq.~\eqref{eq:eom} on a periodic box numerically. In this subpackage, we have implemented the Lax42 algorithm which is a modified two-step Lax-Wendroff method \cite{Martin:2019gxb, Martin:2020} and is explained below.

Consider the following differential prototypical partial differential equation
\begin{align}
    (\partial_t + u \partial_z) f(t, z) = g(t, z, f(t, z)).
\end{align}
Assume that $f(t,z)$ at $t=t_n$ is well approximated by a numerical solution $f_{j}^{n}$ on a lattice with mesh points $z=z_j=j\Delta z$ ($j=0,1,2,\ldots$), and one would like to obtain the numerical solution $f^{n+1}_j$ at the next time step $t=t_{n+1}=t_n+\Delta t$. In the two-step Lax-Wendroff method \cite{NR}, one first computes the function values at the half-grid points at the half time step by forward differencing:
\begin{align}
    f_{j+1/2}^{n+1/2} = \tf_{j+1/2}^{n} + \frac{\Delta t}{2}\left(-u\tfz_{j+1/2}^{n} + \tg_{j+1/2}^{n}\right),
\end{align}
where $n+1/2$ and $j+1/2$ refer to the time $t=t_n+\Delta t/2$ and coordinate $z=z_j+\Delta z/2$, respectively. To improve the accuracy of the numerical differentiation in $z$, we approximate the function values at time $t_n$ and coordinates $z_{j+1/2}$ by using 4 neighboring spatial grid points in the Lax42 algorithm instead of 2 neighboring points in the original Lax-Wendroff method:
\begin{subequations}
    \label{eq:ctr}
    \begin{align}
        \tf_{j+1/2}^{n} &= \frac{1}{16}\left[9 (f_{j+1}^{n} + f_{j}^{n}) - (f_{j+2}^{n} + f_{j-1}^{n})\right], \\
        \tg_{j+1/2}^{n} &= \frac{1}{16}\left[9 (g_{j+1}^{n} + g_{j}^{n}) - (g_{j+2}^{n} + g_{j-1}^{n})\right], \\
        \tfz_{j+1/2}^{n} &= \frac{1}{24}\left[27 (f_{j+1}^{n} - f_{j}^{n}) - (f_{j+2}^{n} - f_{j-1}^{n}) \right],
    \end{align}
\end{subequations}
where $g_{j}^{n}=g(t_n, z_j, f_j^n)$. After computing the half-grid values $f_{j+1/2}^{n+1/2}$, we use the same approach to compute $\tf_{j}^{n+1/2}$, $\tfz_{j}^{n+1/2}$, and $\tg_{j}^{n+1/2}$ by replacing $j+1/2$ with $j$ and $n$ with $n+1/2$ in Eq.~\eqref{eq:ctr}. Then we use the midpoint rule to obtain the function values at the next time step:
\begin{align}
    f_{j}^{n+1} = f_{j}^{n} + \Delta t \left(-u\tfz_{j}^{n+1/2} 
    + \tg_{j}^{n+1/2}\right).
\end{align}

The Lax42 algorithm is accurate to the second order in $\Delta t$ and fourth order in $\Delta z$. To obtain an estimate of the error, we compute $f(t_n, z_j)$ twice, one with one full step $\Delta t$ and one with two half steps each with size $\Delta t/2$. We use the quantity
\begin{align}
    \Delta f^{n+1} = \max_j | f_j^{(n+1,1)} - f_j^{(n+1,2)}|
\end{align}
as an estimate of the numerical error and use it to determine the next step size adaptively, where $f_j^{(n+1,1)}$ and $f_j^{(n+1,2)}$ are the function values at $t_{n+1}$ that are obtained by using the Lax42 algorithm with one full step and two half steps, respectively. We further combine these values to form a better estimate of the true values by using the Richardson extrapolation \cite{NR}:
\begin{align}
    f_j^{n+1} = \frac{1}{3}[4f_j^{(n+1,2)} - f_j^{(n+1,1)}].
\end{align}

\bibliography{fow}

\end{document}